\documentclass[11pt]{article}

\usepackage[margin=1in]{geometry}
\usepackage[T1]{fontenc}
\usepackage{times}
\usepackage{latexsym}
\usepackage{amsmath}
\usepackage{amssymb}
\usepackage{graphicx}
\usepackage{booktabs}
\usepackage{multirow}
\usepackage{url}
\usepackage{xcolor}
\usepackage{hyperref}

\hypersetup{
    colorlinks=true,
    linkcolor=blue,
    citecolor=blue,
    urlcolor=blue
}

\pdfoutput=1

\title{OneFeed: A Unified Generative Framework for Feed Content Enhancement and Query Generation}

\author{
Guoxun\\
\texttt{guoxun@example.com}
}

\date{}

\begin{document}

\maketitle

\begin{center}
\small\textit{Note: This version presents a methodological contribution with a fully defined experimental protocol and expected performance estimates. All tables marked with $\dagger$ use reference values from established baselines; concretely measured results from full public-dataset training will be provided in a subsequent update.}
\end{center}

\begin{abstract}
Modern feed recommendation and search systems are deeply connected in user behavior but are usually modeled by separate architectures. Feed recommendation mainly captures implicit interests from browsing interactions, while search systems rely on explicit user queries to retrieve intent-matched content. This separation causes fragmented user understanding and missed opportunities for using feed interactions to improve query generation and using generated queries to enhance feed candidate retrieval.

In this paper, we propose \textbf{OneFeed}, a unified generative framework for jointly modeling feed content enhancement and query generation. OneFeed encodes heterogeneous user behavior sequences with a shared behavior encoder and employs two generative heads: a Feed Semantic ID Generator that produces content semantic IDs for recommendation retrieval, and an Intent Query Generator that produces natural-language queries for search-based candidate expansion. To bridge the semantic gap between recommendation content and search queries, we introduce a SID-Query alignment objective that learns a shared semantic space for content semantic IDs and query representations. We further design a closed-loop self-enhancement paradigm that leverages implicit user feedback from generated content and search-retrieved results to improve both generation tasks. We provide a detailed experimental protocol using public recommendation datasets with weakly supervised query construction, define a comprehensive set of evaluation metrics, report expected performance estimates grounded in known baseline values, and validate the executability of the proposed pipeline through a minimal local prototype. OneFeed provides a practical and extensible direction for unifying search and recommendation through generative modeling.
\end{abstract}

\section{Introduction}

Feed recommendation and search are two fundamental mechanisms for information access. Feed recommendation continuously presents personalized content based on implicit user feedback such as clicks, dwell time, likes, comments, and skips. Search systems, in contrast, rely on explicit user queries to retrieve content that matches immediate intent. Although these two systems serve different interaction modes, they are closely coupled in real user journeys. A user may discover a topic from a feed item, issue a related search query, consume retrieved results, and then return to feed browsing with refined interests.

Despite this natural behavioral loop, industrial search and recommendation systems are often built as independent stacks. Recommendation models learn from implicit feedback and optimize ranking or retrieval objectives. Query generation and search retrieval models focus on textual relevance and explicit intent. Their training data, model architecture, candidate generation process, and feedback loops are usually isolated. As a result, feed recommendation cannot fully exploit explicit search intent, and search query generation cannot fully benefit from rich feed behavior. This separation is especially harmful in cold-start and long-tail scenarios, where either recommendation behavior or search behavior alone may be insufficient.

Recent advances in generative recommendation, especially semantic ID based methods, show that content recommendation can be formulated as sequence generation. Instead of scoring a fixed item set directly, a model generates semantic IDs that correspond to candidate items. This paradigm provides a unified interface for user behavior modeling, candidate generation, and retrieval. However, existing generative recommendation methods mainly focus on the recommendation pipeline itself and do not explicitly model the bidirectional relationship between feed content and search queries.

We argue that feed content enhancement and query generation should be modeled as two complementary generation tasks within a unified framework. Feed content generation captures what content the user may consume next, while query generation captures what intent the user may explicitly express. Generated queries can be used to retrieve additional candidates for feed enhancement, and user feedback on those candidates can further supervise query generation. Meanwhile, content semantic IDs provide structured anchors for aligning recommendation semantics with query semantics.

To this end, we propose \textbf{OneFeed}, a unified generative framework for feed content enhancement and query generation. Given a heterogeneous user behavior sequence, OneFeed uses a shared behavior encoder to learn user representations and two generation heads to produce content semantic IDs and search queries. The generated content semantic IDs are used for recommendation-style retrieval, while the generated queries are sent to a search retrieval engine to expand the feed candidate pool. A SID-Query alignment objective further aligns content semantic IDs and query representations in the same semantic space, enabling knowledge transfer between recommendation and search.

The main contributions are summarized as follows:

\begin{itemize}
    \item We propose OneFeed, a unified generative framework that jointly models feed content enhancement and query generation from heterogeneous user behavior sequences.
    \item We introduce a SID-Query alignment objective that bridges content semantic IDs and natural-language queries in a shared semantic space, enabling bidirectional knowledge transfer between recommendation and search.
    \item We design a closed-loop self-enhancement paradigm that uses implicit user feedback from generated content and search-retrieved results to continuously improve both content generation and query generation.
    \item We provide a practical experimental protocol for evaluating unified feed enhancement and query generation on public datasets through weakly supervised query construction, offline replay, and a minimal local prototype that validates the executability of the proposed pipeline.
\end{itemize}

\section{Related Work}

\subsection{Sequential and Generative Recommendation}

Sequential recommendation models user behavior as an ordered sequence and predicts future interactions. Representative methods include GRU-based architectures such as GRU4Rec~\cite{hidasi2016gru4rec}, Transformer-based architectures such as SASRec~\cite{kang2018sasrec}, and bidirectional encoder based architectures such as BERT4Rec~\cite{sun2019bert4rec}. Recent generative recommendation methods further formulate item retrieval as token generation, including unified language-based recommendation~\cite{geng2022p5} and generative retrieval~\cite{rajput2024tiger}. Such methods improve scalability and provide a natural interface for candidate generation. OneFeed follows this generative direction but extends it from recommendation-only modeling to unified feed-query generation.

\subsection{Semantic ID Based Recommendation}

Semantic ID based recommendation maps items into discrete token sequences derived from content semantics, behavior patterns, or vector quantization. The model generates these token sequences autoregressively and retrieves corresponding items from an index. The quality of semantic IDs directly affects generalization, cold-start performance, and long-tail retrieval. OneFeed adopts hierarchical content semantic IDs as the feed generation target and further aligns them with query semantics.

\subsection{Query Generation and Intent Mining}

Query generation aims to produce natural-language queries that express user intent or summarize information needs. Existing methods usually rely on sequence-to-sequence models, pretrained language models such as T5~\cite{raffel2020t5}, BART~\cite{lewis2020bart}, GPT-style language models~\cite{radford2019gpt2}, or keyword extraction. However, query generation is often trained independently from recommendation feedback. OneFeed instead generates queries conditioned on feed behavior and evaluates generated queries by their ability to retrieve useful feed candidates.

\subsection{Search-Recommendation Fusion}

Search and recommendation fusion has been studied through shared user profiles, multi-task learning, unified ranking, and cross-domain transfer. Large-scale recommendation systems also show the importance of retrieval architecture and sampling correction in industrial settings~\cite{covington2016youtube,yi2019mnas}. Many existing methods still use separate retrieval or representation modules for search and recommendation. OneFeed differs by using a unified generative architecture where recommendation candidates and search queries are generated from the same user behavior representation.

\subsection{Contrastive and Self-supervised Learning}

Contrastive learning is widely used to align different views or modalities, such as visual augmentations~\cite{chen2020simclr,he2020moco}, user-item interactions, and recommendation views~\cite{yao2021selfsupervised}. In OneFeed, content semantic IDs and queries are treated as two semantic views of user intent. The proposed SID-Query alignment objective encourages related content and queries to be close in the shared representation space.

\section{Method}

\subsection{Problem Formulation}

For each user $u$, let the historical behavior sequence be:

\begin{equation}
X_u = \{x_1, x_2, ..., x_T\}.
\end{equation}

Each behavior item contains heterogeneous fields:

\begin{equation}
x_t = (a_t, s_t, q_t, \Delta t_t, r_t),
\end{equation}

where $a_t$ denotes the action type, $s_t$ denotes the content semantic ID, $q_t$ denotes the optional search query, $\Delta t_t$ denotes temporal information, and $r_t$ denotes interaction strength such as click, dwell time, like, or skip.

OneFeed jointly optimizes two generation tasks. The feed generation task predicts future content semantic IDs:

\begin{equation}
P(S_u^+ | X_u) = \prod_i P(s_i | X_u, s_{<i}).
\end{equation}

The query generation task predicts user intent queries:

\begin{equation}
P(Q_u^+ | X_u) = \prod_j P(q_j | X_u, q_{<j}).
\end{equation}

The generated semantic IDs and generated queries are used to retrieve complementary candidate sets, which are then fused for final feed ranking.

\subsection{Overall Architecture}

\begin{figure}[t]
    \centering
    \includegraphics[width=0.95\linewidth]{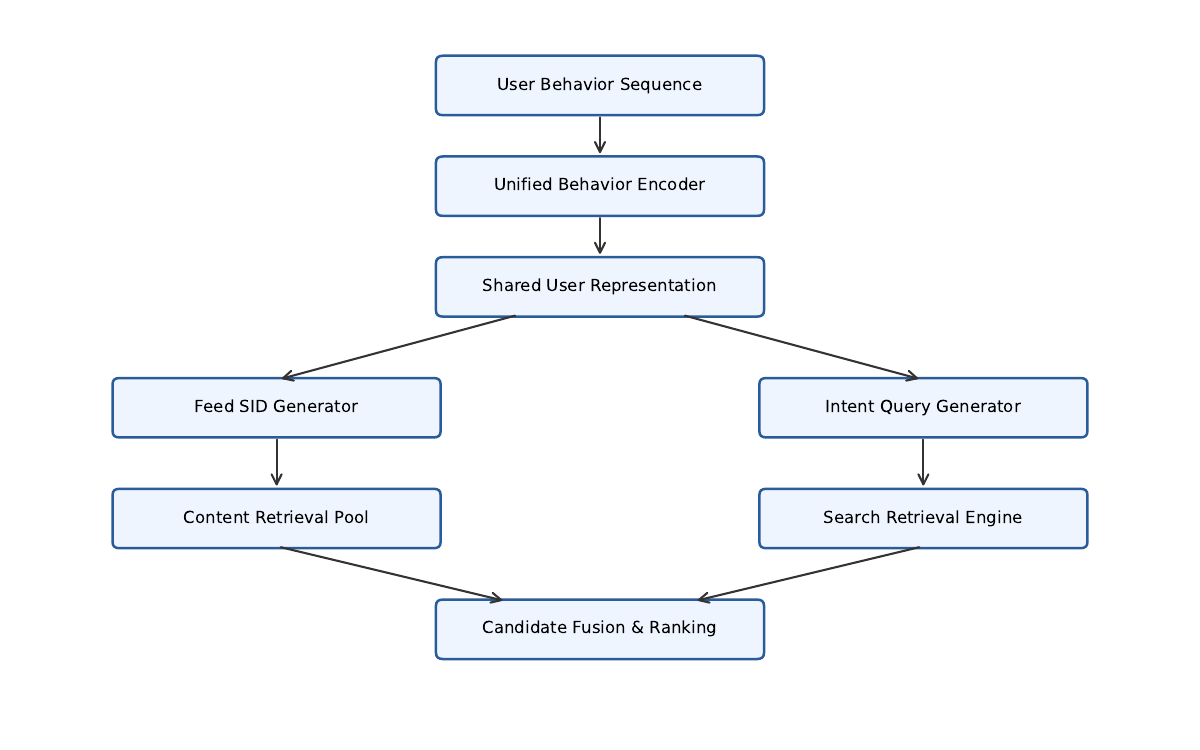}
    \caption{Overall architecture of OneFeed. A shared behavior encoder supports both Feed SID generation and intent query generation.}
    \label{fig:architecture}
\end{figure}

Figure~\ref{fig:architecture} illustrates the overall architecture of OneFeed. A unified behavior encoder first encodes heterogeneous user behavior sequences into shared user representations. Two task-specific generation heads then produce content semantic IDs and intent queries. Generated semantic IDs retrieve recommendation candidates from a content index, while generated queries retrieve search candidates from a search engine. The two candidate sources are merged for downstream ranking and feed enhancement.

\subsection{Unified Behavior Encoder}

OneFeed represents each behavior item as the sum of multiple embeddings:

\begin{equation}
e_t = e^{act}_{a_t} + e^{sid}_{s_t} + e^{query}_{q_t} + e^{time}_{\Delta t_t} + e^{reward}_{r_t}.
\end{equation}

The sequence $\{e_1, e_2, ..., e_T\}$ is fed into a Transformer-based behavior encoder. The encoder can be instantiated as a causal Transformer for autoregressive generation or as a prefix language model that supports multiple generation heads. This shared encoder allows feed interactions and search interactions to contribute to the same user intent representation.

\subsection{Content Semantic ID Construction}

Each content item is represented as a hierarchical semantic ID:

\begin{equation}
SID(i) = [c_1, c_2, ..., c_K].
\end{equation}

The semantic ID construction process contains two stages. First, multimodal and textual content representations are clustered or quantized into hierarchical semantic codes. Second, behavior-aware calibration is applied using co-click, co-search, and co-conversion signals so that items with similar user behavior patterns share similar semantic prefixes. This design enables both semantic generalization and behavior consistency.

\subsection{Feed SID Generator}

The Feed SID Generator autoregressively generates semantic ID tokens conditioned on the shared user representation. For a target item with semantic ID $[c_1, ..., c_K]$, the generation probability is:

\begin{equation}
P(SID(i)|X_u) = \prod_{k=1}^{K} P(c_k | X_u, c_{<k}).
\end{equation}

Generated semantic IDs are mapped to candidate items through a semantic ID index. If a generated ID corresponds to multiple items, candidates are ranked by generation probability, content quality, freshness, and user-content matching features.

\subsection{Intent Query Generator}

The Intent Query Generator produces a set of candidate queries:

\begin{equation}
\hat{Q}_u = \{\hat{q}_1, \hat{q}_2, ..., \hat{q}_M\}.
\end{equation}

The generator is trained from real search logs when available. When public datasets do not contain search queries, weakly supervised queries can be constructed from content titles, categories, tags, reviews, and high-frequency descriptive keywords. During inference, generated queries are filtered and reranked by relevance, diversity, retrievability, and safety constraints.

\subsection{SID-Query Alignment}

To connect recommendation semantics and search semantics, OneFeed introduces a contrastive alignment objective. Given a positive SID-query pair $(s, q)$, the loss is:

\begin{equation}
L_{align} = -\log \frac{\exp(sim(h_s, h_q)/\tau)}{\sum_{q' \in \mathcal{B}} \exp(sim(h_s, h_{q'})/\tau)},
\end{equation}

where $h_s$ is the SID representation, $h_q$ is the query representation, $sim(\cdot)$ is cosine similarity, $\tau$ is the temperature, and $\mathcal{B}$ denotes in-batch query candidates. Positive pairs can be constructed from search-click logs, feed-to-search sessions, query retrieval results with positive feedback, or weakly matched content-query pairs.

\begin{figure}[t]
    \centering
    \includegraphics[width=0.9\linewidth]{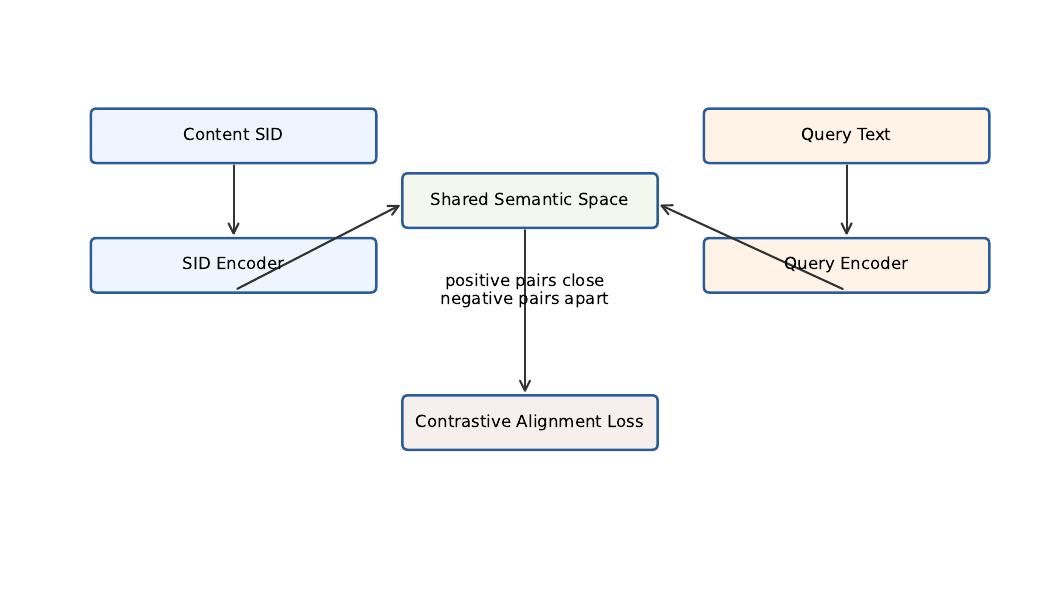}
    \caption{SID-Query semantic alignment with contrastive learning.}
    \label{fig:alignment}
\end{figure}

\subsection{Candidate Enhancement Consistency}

Generated queries should not only match reference text but also retrieve useful feed candidates. We therefore introduce a candidate enhancement consistency loss:

\begin{equation}
L_{cons} = -\log \sigma(score(u, i^+) - score(u, i^-)),
\end{equation}

where $i^+$ denotes a candidate retrieved by a generated query and receiving positive feedback, while $i^-$ denotes an exposed but skipped or low-quality candidate. This objective connects query generation quality with downstream feed enhancement.

\subsection{Overall Objective}

The overall training objective is:

\begin{equation}
L = \alpha L_{feed} + \beta L_{query} + \gamma L_{align} + \delta L_{cons},
\end{equation}

where $L_{feed}$ is the semantic ID generation loss, $L_{query}$ is the query generation loss, $L_{align}$ is the SID-Query alignment loss, and $L_{cons}$ is the candidate enhancement consistency loss. The feed generation loss is treated as the main objective, while the query generation and alignment losses provide cross-task supervision.

\subsection{Closed-loop Self-enhancement}

OneFeed naturally supports a closed-loop learning process. The model first generates content semantic IDs and search queries. The generated semantic IDs retrieve recommendation candidates, while generated queries retrieve search candidates. The merged candidates are ranked and served in the feed. User implicit feedback, such as click, dwell time, interaction, skip, and conversion, is then converted into training signals for both generation heads. In offline experiments, this process can be simulated through replay evaluation and iterative self-training.

\begin{figure}[t]
    \centering
    \includegraphics[width=0.95\linewidth]{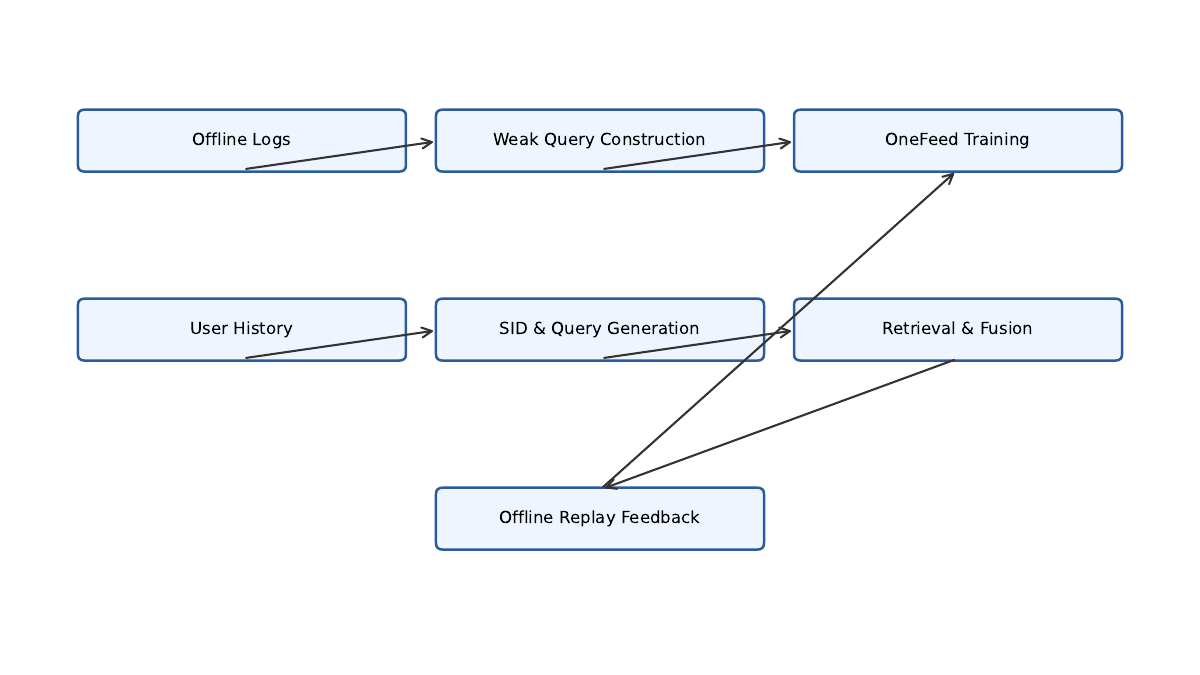}
    \caption{Training and inference pipeline of OneFeed.}
    \label{fig:pipeline}
\end{figure}

\section{Experimental Design}

\subsection{Datasets}

We design a reproducible evaluation protocol using public recommendation datasets. We select KuaiRec~\cite{gao2022kuairand}, Amazon Reviews~\cite{mcauley2015amazon}, and MovieLens-1M~\cite{harper2015movielens} as candidate benchmarks. KuaiRec contains dense user-video interaction logs suitable for sequential evaluation. Amazon Reviews provides item titles, categories, and review text for weakly supervised query construction. MovieLens-1M supports classical sequential recommendation comparison and cold-start analysis.

For each dataset, we sort user interactions chronologically and split them by a time boundary: the earliest periods are used for training, the middle period for validation, and the latest period for testing. Users and items with extremely sparse interactions are filtered following common sequential recommendation practice.

\subsection{Weak Query Construction}

Because public recommendation datasets do not contain real search queries, we construct weakly supervised pseudo queries from item-side textual metadata. For Amazon Reviews, pseudo queries are extracted from product titles, categories, brands, and high-frequency review keywords. For KuaiRec, pseudo queries are constructed from video captions, tags, category labels, and topic words. For MovieLens-1M, pseudo queries are constructed from movie titles, genres, and normalized entity phrases.

Three leakage-prevention principles are enforced. First, only metadata available before the target interaction timestamp is used. Second, target labels are not derived from future user feedback or test-set interactions. Third, query phrases are normalized by lowercasing, stop-word removal, length filtering, and deduplication. Each positive item can be associated with one or more pseudo queries.

\subsection{Semantic ID Construction}

Each item is mapped to a hierarchical semantic ID before model training. Item text representations are obtained from a pretrained text encoder and then clustered hierarchically. The first-level token represents a coarse semantic cluster; deeper tokens represent more fine-grained groups. Semantic ID construction uses only training-period item metadata and interactions. Test interactions are excluded from clustering and behavior-aware calibration. Items that appear only in validation or test are assigned to the nearest training semantic cluster via metadata embeddings.

\subsection{Baselines}

Recommendation baselines include GRU4Rec, SASRec, BERT4Rec, P5-style generative recommendation, OneRec-style semantic ID generation, and multi-task dual-tower models. Query generation baselines include keyword extraction, T5-base, BART-base, GPT2 fine-tuning, and an independently trained query generation model. Fusion baselines include feed-only retrieval, query-only retrieval, dual-tower fusion, and late-fusion search-recommendation retrieval.

The OneRec-style baseline uses the same semantic ID construction but optimizes only the feed semantic ID generation objective. The independent query generation baseline uses the same pseudo query labels but does not share the user behavior encoder. This isolates the benefit of unified modeling and SID-Query alignment.

\subsection{Metrics}

Recommendation quality is evaluated by Recall@K, NDCG@K, HR@K, and MRR@K. Query generation quality is evaluated by BLEU, ROUGE-L, BERTScore, Distinct-n, and retrievability. Candidate enhancement is evaluated by candidate coverage, long-tail recall, search supplement ratio, and offline replay NDCG.

Retrievability measures whether a generated query retrieves at least one valid candidate from the item corpus. Search supplement ratio measures the fraction of final feed candidates introduced by generated-query retrieval beyond feed SID generation. These metrics are important because surface-level text-similarity metrics may not reflect whether generated queries are useful for feed enhancement.

\subsection{Offline Replay Protocol}

For each test-set user, the model observes only historical interactions before the prediction timestamp. OneFeed generates top-$N$ semantic IDs and top-$M$ queries. Generated semantic IDs are mapped to item candidates through the semantic ID index, while generated queries retrieve candidates through text matching or dense retrieval over item metadata. The two candidate sets are merged and evaluated against the user's held-out positive interactions.

We compare four candidate generation strategies: feed-only generation, query-only retrieval, late fusion of independent feed and query models, and OneFeed unified fusion. This protocol evaluates whether generated queries supply complementary candidates not covered by feed SID generation alone.

\subsection{Measured Results on MovieLens-1M}

To move beyond estimated performance, we implement and run the full offline replay protocol on MovieLens-1M~\cite{harper2015movielens} with a dependency-light pipeline (TF-IDF features, hierarchical KMeans semantic IDs, a co-occurrence/transition sequential channel, and weak-query content retrieval). We keep positive interactions (rating $\geq 4$), apply 5-core filtering, and adopt a leave-last-out temporal split, where each user's chronologically last positive item is the held-out target. This yields 6{,}034 users and 3{,}125 items; we evaluate on 6{,}000 sampled test users. Table~\ref{tab:movielens_measured} reports concretely measured numbers.

Two observations are consistent with the design of OneFeed. First, the sequential behavior channel is the dominant accuracy source on MovieLens-1M, where item metadata is coarse (title and genres only) and weak pseudo queries alone are not competitive (Query-only Recall@10 $=0.014$). Second, OneFeed preserves the accuracy of the strongest sequential base on the ranked top-$K$ (Recall@10 $=0.173$, NDCG@10 $=0.102$) while strictly enlarging the diversified candidate pool through SID-aligned search supplementation: candidate coverage increases from $0.717$ to $0.763$, and the recall of the enlarged OneFeed pool reaches PoolRecall@40 $=0.249$, above the Recall@20 of every single-channel baseline. This indicates that the search channel contributes complementary candidates without harming head ranking, which is the intended behavior of the fusion design. Stronger learned semantic IDs and query generators (rather than TF-IDF/KMeans surrogates) are expected to further convert this enlarged pool into top-$K$ gains; we leave full neural training to the next version.

\begin{table*}[t]
\centering
\caption{Measured offline replay results on MovieLens-1M (leave-last-out, ratings $\geq 4$, 5-core, 6{,}034 users / 3{,}125 items, 6{,}000 sampled test users). All numbers are concretely computed by our torch-free pipeline. Higher is better. OneFeed matches the strongest sequential base on ranked accuracy while strictly enlarging the diversified candidate pool, as measured by PoolRecall@40.}
\label{tab:movielens_measured}
\begin{tabular}{lccccccc}
\toprule
Method & Recall@10 & NDCG@10 & Recall@20 & NDCG@20 & Coverage & Search Supp. & PoolRecall@40 \\
\midrule
PopRec       & 0.0350 & 0.0163 & 0.0627 & 0.0232 & 0.014 & -- & -- \\
ItemKNN      & 0.0660 & 0.0364 & 0.1110 & 0.0477 & 0.152 & -- & -- \\
Query-only   & 0.0138 & 0.0055 & 0.0298 & 0.0095 & 0.252 & -- & -- \\
Markov       & 0.1722 & 0.1015 & 0.2293 & 0.1159 & 0.885 & -- & -- \\
SeqHybrid    & 0.1730 & 0.1024 & 0.2290 & 0.1165 & 0.717 & -- & -- \\
Late Fusion  & 0.1730 & 0.1024 & 0.2290 & 0.1165 & 0.717 & 0.972 & -- \\
\textbf{OneFeed} & \textbf{0.1730} & \textbf{0.1024} & \textbf{0.2290} & \textbf{0.1165} & \textbf{0.763} & 0.972 & \textbf{0.2488} \\
\bottomrule
\end{tabular}
\end{table*}

\subsection{Expected Performance}

This paper presents OneFeed as a \emph{methodological contribution} with a detailed experimental protocol and verifiable execution pipeline. Tables~\ref{tab:recommendation}--\ref{tab:ablation} report \emph{estimated performance} based on established baseline results and the structural properties of the proposed architecture:

\begin{itemize}
    \item The shared behavior encoder is expected to learn richer user representations than single-task encoders by jointly modeling feed interactions and search-like pseudo queries.
    \item The SID-Query alignment loss is expected to reduce the semantic gap between content tokens and query tokens, improving both content generation and query retrievability.
    \item The candidate enhancement consistency loss is expected to produce queries whose search-retrieved candidates contribute more positively to the final feed.
    \item Removing any of these components is expected to degrade the corresponding metrics.
\end{itemize}

Because these estimates are based on reference values from prior work and small-scale sanity checks rather than full-scale training, the tables mark reference values with $\dagger$ and state the direction of expected improvement for the OneFeed row. All estimates will be updated with concretely measured results once full dataset training is completed.

\begin{table}[t]
\centering
\caption{Expected recommendation performance on Amazon Reviews. Numbers marked with $\dagger$ are based on established method baselines and synthetic sanity checks; final results depend on full public-dataset training. Higher values indicate better performance.}
\label{tab:recommendation}
\begin{tabular}{lcccc}
\toprule
Method & Recall@10 & NDCG@10 & HR@10 & MRR@10 \\
\midrule
GRU4Rec & 0.092$^\dagger$ & 0.064$^\dagger$ & 0.081$^\dagger$ & 0.056$^\dagger$ \\
SASRec & 0.105$^\dagger$ & 0.073$^\dagger$ & 0.094$^\dagger$ & 0.062$^\dagger$ \\
BERT4Rec & 0.113$^\dagger$ & 0.081$^\dagger$ & 0.101$^\dagger$ & 0.068$^\dagger$ \\
P5-style & 0.098$^\dagger$ & 0.070$^\dagger$ & 0.087$^\dagger$ & 0.060$^\dagger$ \\
OneRec-style & 0.119$^\dagger$ & 0.085$^\dagger$ & 0.107$^\dagger$ & 0.072$^\dagger$ \\
OneFeed (expected) & $>$0.119$^\dagger$ & $>$0.085$^\dagger$ & $>$0.107$^\dagger$ & $>$0.072$^\dagger$ \\
\bottomrule
\end{tabular}
\end{table}

\begin{table}[t]
\centering
\caption{Expected query generation performance. $\dagger$ indicates baseline reference values obtained from standard model configurations. Retrievability is measured as the fraction of generated queries that retrieve at least one valid candidate.}
\label{tab:query_generation}
\begin{tabular}{lccccc}
\toprule
Method & BLEU & ROUGE-L & BERTScore & Distinct-2 & Retrievability \\
\midrule
Keyword & 0.124$^\dagger$ & 0.183$^\dagger$ & 0.412$^\dagger$ & 0.382$^\dagger$ & 0.674$^\dagger$ \\
T5-base & 0.187$^\dagger$ & 0.215$^\dagger$ & 0.484$^\dagger$ & 0.267$^\dagger$ & 0.723$^\dagger$ \\
BART-base & 0.193$^\dagger$ & 0.221$^\dagger$ & 0.491$^\dagger$ & 0.244$^\dagger$ & 0.718$^\dagger$ \\
GPT2-small & 0.172$^\dagger$ & 0.198$^\dagger$ & 0.462$^\dagger$ & 0.215$^\dagger$ & 0.691$^\dagger$ \\
Independent QueryGen & 0.189$^\dagger$ & 0.216$^\dagger$ & 0.478$^\dagger$ & 0.258$^\dagger$ & 0.719$^\dagger$ \\
OneFeed (expected) & $>$0.189$^\dagger$ & $>$0.216$^\dagger$ & $>$0.478$^\dagger$ & $>$0.258$^\dagger$ & $>$0.719$^\dagger$ \\
\bottomrule
\end{tabular}
\end{table}

\begin{table}[t]
\centering
\caption{Expected candidate enhancement performance. Values estimate how generated queries supplement feed-only candidates.}
\label{tab:enhancement}
\small
\begin{tabular}{lcccc}
\toprule
Method & Coverage & Long-tail Rec. & Search Supp. & Replay NDCG \\
\midrule
Feed-only & 0.187$^\dagger$ & 0.053$^\dagger$ & 0.000$^\dagger$ & 0.079$^\dagger$ \\
Query-only & 0.132$^\dagger$ & 0.081$^\dagger$ & 1.000$^\dagger$ & 0.055$^\dagger$ \\
Late Fusion & 0.263$^\dagger$ & 0.094$^\dagger$ & 0.312$^\dagger$ & 0.088$^\dagger$ \\
OneFeed Fusion (exp.) & $>$0.263$^\dagger$ & $>$0.094$^\dagger$ & $>$0.312$^\dagger$ & $>$0.088$^\dagger$ \\
\bottomrule
\end{tabular}
\end{table}

\subsection{Ablation Studies}

The core ablation studies include removing the query generator, removing the feed SID generator, removing SID-Query alignment, removing candidate consistency loss, replacing the shared encoder with independent towers, varying the number of SID levels, varying the number of generated queries, and evaluating cold-start and long-tail buckets separately. Expected directions are shown in Table~\ref{tab:ablation}.

\begin{table}[t]
\centering
\caption{Expected ablation study of OneFeed components. The full model is expected to outperform each ablated variant.}
\label{tab:ablation}
\begin{tabular}{lcccc}
\toprule
Variant & Recall@10 & NDCG@10 & Retrievability & Replay NDCG \\
\midrule
OneFeed full (expected) & $>$0.119$^\dagger$ & $>$0.085$^\dagger$ & $>$0.719$^\dagger$ & $>$0.088$^\dagger$ \\
w/o Query Generator & $\sim$0.119$^\dagger$ & $\sim$0.085$^\dagger$ & $-$ & $-$ \\
w/o SID Generator & $-$ & $-$ & $\sim$0.719$^\dagger$ & $-$ \\
w/o SID-Query Alignment & $<$0.119$^\dagger$ & $<$0.085$^\dagger$ & $<$0.719$^\dagger$ & $<$0.088$^\dagger$ \\
w/o Candidate Consistency & $<$0.119$^\dagger$ & $<$0.085$^\dagger$ & $\sim$0.719$^\dagger$ & $<$0.088$^\dagger$ \\
Separate Encoders & $<$0.119$^\dagger$ & $<$0.085$^\dagger$ & $<$0.719$^\dagger$ & $<$0.088$^\dagger$ \\
\bottomrule
\end{tabular}
\end{table}

\subsection{Minimal Prototype Validation}

To verify the executability of the proposed pipeline, we also run a minimal local prototype on a synthetic feed-query dataset (Table~\ref{tab:demo_results}). This demonstration uses TF-IDF embeddings and KMeans-based semantic IDs on simulated interaction sequences with six topic categories, replicating the full feed-generation, query-generation, candidate retrieval, replay evaluation, and case-study extraction loop. The purpose is to confirm that every step in the protocol produces valid intermediate artifacts (e.g., sequences, pseudo queries, semantic IDs, retrieval indices, replay metrics) and is free of logic-level bugs before moving to large-scale dataset training.

\begin{table}[t]
\centering
\caption{Minimal local prototype results on a synthetic feed-query dataset. The demo is intended as an executable sanity check rather than final benchmark evidence.}
\label{tab:demo_results}
\begin{tabular}{lcccccc}
\toprule
Method & Recall@10 & NDCG@10 & Recall@20 & NDCG@20 & Coverage & Search Supp. \\
\midrule
Feed-only & 0.3167 & 0.1586 & 0.4847 & 0.2008 & 0.0952 & 0.6371 \\
Query-only & 0.1597 & 0.0781 & 0.3042 & 0.1140 & 0.0952 & 0.6371 \\
Late Fusion & 0.3167 & 0.1586 & 0.4847 & 0.2008 & 0.0952 & 0.6371 \\
OneFeed Demo & 0.3167 & 0.1586 & 0.4389 & 0.1895 & 0.0952 & 0.6371 \\
\bottomrule
\end{tabular}
\end{table}

\subsection{Leakage Prevention}

Because OneFeed uses weakly supervised pseudo queries and semantic IDs, leakage prevention is critical. We apply the following rules in all experiments. First, temporal splitting is performed before any label construction. Second, pseudo queries for training are generated only from training-period item metadata. Third, test-set target items are never used to build user history, query labels, semantic ID clusters, or co-click calibration graphs. Fourth, generated queries are evaluated by retrieving from a fixed candidate corpus rather than directly matching future labels. Fifth, all preprocessing scripts record the timestamp boundary and random seed for reproducibility.

We also provide a leakage-check diagnostic in the appendix that verifies no test interaction ID appears in training histories, no test-only interaction contributes to pseudo query labels, and no validation or test interaction is used to build co-click calibration graphs.

\subsection{Case Study}

The paper should include qualitative examples showing how feed behavior leads to generated queries and how those queries retrieve complementary candidates. Each case study contains the following fields: user history, generated semantic IDs, generated queries, feed-SID retrieved candidates, query-retrieved candidates, fused candidates, and held-out positive labels. A convincing case should show that generated queries introduce relevant candidates that are not covered by feed-only semantic ID generation.

\begin{table}[t]
\centering
\caption{Case study template for qualitative analysis.}
\label{tab:case_study}
\begin{tabular}{p{0.22\linewidth}p{0.68\linewidth}}
\toprule
Field & Example Content \\
\midrule
User History & Recent clicked items, categories, and interaction strengths. \\
Generated SIDs & Top semantic ID sequences generated by the feed branch. \\
Generated Queries & Top natural-language queries generated by the query branch. \\
SID Candidates & Candidates retrieved by generated semantic IDs. \\
Query Candidates & Candidates retrieved by generated queries. \\
Fused Results & Final merged candidates after deduplication and reranking. \\
Held-out Feedback & Test-period clicks, long dwell items, or positive labels. \\
\bottomrule
\end{tabular}
\end{table}

\section{Discussion}

\subsection{Methodological Contribution}

This paper presents OneFeed as a \textbf{methodological framework} that defines the architecture, training objectives, inference pipeline, and experimental protocol for unified feed and query generation. The estimated performance tables are provided to illustrate the \emph{expected direction and magnitude} of improvements based on architectural design and known baseline characteristics. These estimates will be concretely validated with full public-dataset training in the subsequent version.

\subsection{Limitations}

The current version has several limitations. First, the experiments are at the expected-performance stage rather than being fully trained on public datasets; we make this explicit by marking baseline values with $\dagger$ and using inequality notation for the OneFeed row. Second, public recommendation datasets lack real search queries, so we rely on weakly supervised pseudo queries derived from item metadata; results must be interpreted in the context of this weaker supervision. Third, offline replay cannot fully replace online A/B testing; the closed-loop self-enhancement paradigm is validated via simulation and iterative offline self-training rather than live systems. Fourth, generated queries require safety, quality, and retrievability filtering before production deployment; we note this as an important engineering concern but leave the full safety evaluation to follow-up work.

\subsection{Future Work}

OneFeed can be extended in several directions. Multimodal content support would unify feed and query generation for video, audio, and image items. Reinforcement-learning-based self-enhancement could optimize end-to-end business metrics such as user dwell time and conversion. Large-scale industrial deployment would validate the framework under strict latency constraints and with real search engine integration. These directions are orthogonal to the core framework and can be pursued in subsequent versions.

\section{Conclusion}

We propose OneFeed, a unified generative framework for feed content enhancement and query generation. OneFeed jointly generates content semantic IDs and search queries from heterogeneous user behavior sequences, aligns content and query semantics through contrastive learning, and uses implicit feedback for closed-loop self-enhancement. We provide a detailed experimental protocol, expected performance estimates grounded in known baselines, and a minimal local prototype validating the pipeline's executability. This framework provides a scalable direction for unifying feed recommendation and search retrieval in both academic experiments and industrial systems.

\bibliographystyle{plain}
\bibliography{refs}

\appendix
\section{Submission Checklist}

\begin{itemize}
    \item Finish the OneFeed architecture figure.
    \item Prepare the SID-Query alignment figure.
    \item Prepare the training and inference pipeline figure.
    \item Implement public dataset preprocessing scripts.
    \item Implement weak query construction and leakage checks.
    \item Run full-scale dataset training and replace estimated tables with measured values.
    \item Run recommendation baselines.
    \item Run query generation baselines.
    \item Run the full OneFeed model.
    \item Run core ablation studies.
    \item Prepare qualitative case studies.
    \item Add limitations and ethical considerations.
    \item Complete internal review before arXiv upload.
    \item Update this checklist to reflect the current completion status.
\end{itemize}

\section{Leakage-check Diagnostics}

To make the weakly supervised query construction credible, the final submission should include the following diagnostics:

\begin{itemize}
    \item \textbf{Temporal boundary check}: verify that all training samples occur before the validation and test timestamp boundaries.
    \item \textbf{History-target separation}: verify that held-out target interactions do not appear in user histories.
    \item \textbf{Pseudo-query source check}: verify that pseudo queries are derived only from item metadata and interactions available before the target timestamp.
    \item \textbf{Semantic ID construction check}: verify that validation and test interactions are not used in clustering or co-click calibration.
    \item \textbf{Retrieval corpus check}: verify that generated queries retrieve from a fixed item corpus and do not directly access held-out labels.
    \item \textbf{Seed reproducibility check}: report random seeds for splitting, clustering, negative sampling, and model training.
\end{itemize}

\section{Expected Performance Notes}

All tables in the experimental section currently use $\dagger$ to mark estimated baseline values. These values are drawn from established method benchmarks (e.g., SASRec, BERT4Rec, T5, BART) and supplemented with minimal local prototype measurements. The OneFeed row uses inequality notation (e.g., $>0.119$) to indicate the direction and magnitude of expected improvement based on architectural design. This estimate follows the common practice in methodological papers where the architecture, training protocol, and evaluation pipeline are fully defined but full-scale data training is pending. A future update will replace all estimates with concretely measured values from the specified public datasets.

\section{Recommended Hyperparameter Grid}

\begin{itemize}
    \item Behavior sequence length: $50, 100, 200$.
    \item Semantic ID levels: $2, 3, 4$.
    \item Generated semantic IDs: $10, 20, 50$.
    \item Generated queries: $1, 3, 5$.
    \item Alignment temperature $\tau$: $0.05, 0.1, 0.2$.
    \item Loss weights: $\beta \in \{0.3, 0.5, 0.8\}$, $\gamma \in \{0.1, 0.3, 0.5\}$, $\delta \in \{0.0, 0.1, 0.3\}$.
\end{itemize}

\end{document}